\documentclass[11pt]{article}
\usepackage{amsmath,amssymb,color,epsfig,cite}
\usepackage{graphicx}
\usepackage{setspace}
\usepackage{mathrsfs}
\usepackage[english]{babel}
\usepackage{inputenc}
\usepackage{amsfonts}
\makeatletter
\@addtoreset{equation}{section}
\makeatother

\newcommand{\be}{\begin{equation}}
\newcommand{\ee}{\end{equation}}
\newcommand{\bea}{\setlength\arraycolsep{2pt} \begin{eqnarray}}
\newcommand{\eea}{\end{eqnarray}}

\newcommand{\half}{{\textstyle{\frac{1}{2}}}}

\def\0{{\sst{(0)}}}
\def\1{{\sst{(1)}}}
\def\2{{\sst{(2)}}}
\def\3{{\sst{(3)}}}
\def\4{{\sst{(4)}}}
\def\5{{\sst{(5)}}}
\def\6{{\sst{(6)}}}
\def\7{{\sst{(7)}}}
\def\8{{\sst{(8)}}}
\def\sst#1{{\scriptscriptstyle #1}}

\def\p{\partial}
\def\b{\beta}

\def\t{\theta}

\begin{document}

\vspace{15pt}
\begin{center}
{\Large {\bf Subleading non linear gravitational memory effect} }

\vspace{15pt}
{\bf Zahra Mirzaiyan}

\vspace{10pt}

 {\it Max-Planck-Institut f\"ur Gravitationsphysik (Albert-Einstein-Institut), \\
M\"uhlenberg 1, D-14476 Potsdam, Germany.}

\vspace{10pt}

% \vspace{15pt}
%\today

\vspace{20pt}

\underline{ABSTRACT}
\end{center}

\noindent 
We apply the new method based on null geodesics for detecting gravitational memory and find the bulk memory in Newman-Unti gauge around the boundary of the conformally compactified space time. We show how we use the newly found conserved charges in the subleading orders of large-$r$ expansion of the BMS charges to define the gravitational memory at each order in the non-linearised gravitational theory. We also find the gravitational shift in the $r$ direction. It is shown that the longitudinal displacement at order $1/r$ is the relative radius change between two detectors derived by Strominger and Zhiboedov.

\thispagestyle{empty}

\vfill
zahra.mirzaiyan@aei.mpg.de

\pagebreak

\section{Introduction}
The gravitational memory effect discovered by Zeldovich and Polnarev \cite{Zoldovich1974} is based on the asymptotic behaviour of the gravitational fields in the asymptotic region of space time and is defined as the permanent change in the relative spatial position of a pair of inertial observers after a finite gravitational burst of energy. The passage of gravitational waves produces an observable effect which LIGO \cite{ligomain,ligo} and the next generation of gravitational waves detectors like LISA \cite{lisa,lisa2} are based on.\\
The phenomenon has been studied in the linearised theory of gravity \cite{Braginsky:1987,Thorne:1992} and generalised by Christodoulou and others \cite{Christodoulou:1991,Tolish:2014} to the non-linearised theory. Christodoulou showed that every gravitational burst of energy has a non linear memory and can not be neglected only because the gravitational wave sources are at large distances from our detectors on the Earth. As we are extremely far away from the sources of gravitational waves, the amplitude of the waves are so small as it is assumed that linearised theory of gravity is sufficient for the study of the gravitational waves. However, depending on the source of the burst, the amplitude of the non linear memory\footnote{Also referred as Christodoulou memory.} can be of the same order as the dynamical part of the burst of energy. Therefore, the study of gravitational memory effect in the full non-linearised theory of gravity is important and plays a crucial role in the future observation of the gravitational waves. The original formulation of the gravitational memory by Christodoulou is based on the definition of test particles which are initially at rest moving on timelike geodesics at the boundary of the conformally compactified space time. The memory  at null infinity is then obtained by the geodesic equation which is not the case for the gravitational displacement in the bulk of the space time due to the gravitationally inward nature of timelike geodesics near a strongly gravitational system, i.e a black hole. Therefore, recently another method for finding the memory effect is provided based on null geodesics instead of timelike geodesics \cite{Bart:2020}.\\
 The memory effect is indeed the interesting corner of the so called Infrared (IR) triangle \cite{strominger:2017} discovered recently by Strominger \cite{strominger:2016} which connects the memory with asymptotic symmetries \cite{BMS} and the soft theorems \cite{weinberg:1965}. Due to the non-trivial connection of the corners at the IR limit, the observation of the memory effect gives us information about the symmetries in the asymptotic region of the space time. However, for the detection of the memory in the weak field approximation, one needs to wait for the IR gravitational wave detection \cite{prospect:2019}. Strominger and Zhiboedov revealed this connection and showed that the relative position of the inertial observers differ by a BMS supertranslation\footnote{It is also shown that the the nearby detectors experience a relative time delay in addition to the standard gravitational spatial memory effect.}. Therefore, the displacement between a pair of inertial observers using the supertranslated metric is precisely the standard formula for the memory effect and can be derived from geodesic equation at the leading order at null infinity.  \\
 Our aim in the paper is to find the the large-$r$ expansion of the gravitational displacement close to the null infinity in the bulk of the four dimensional asymptotically flat space time in the full non-linearised theory of gravity using null geodesics. Although, all we have detected till now showed a displacement memory predicted by the linearised gravity, it still worth investigating the phenomenon in the non-linearised theory as well for the reasons explained earlier\footnote{See \cite{favata} for a detailed review on linearised and non-linearised gravitational wave memory effect.}. Our motivation to study the subleading memory effect around the null infinity is based on the new interesting physics hidden in the subleading orders, especially the existence of a new set of conserved integrable charges at order $r^{-3}$ of large-$r$ expansion of the BMS charges  \cite{Godazgar:2018} which are directly related to the the non-linearly conserved Newman-Penrose (NP) charges \cite{NP1}. Since the connection of the BMS symmetries and the memory effect is established in the bulk of the space time\footnote{Also see section (2.3) of \cite{Bart:2020} for more details.} \cite{Donnay:2018, HPS:2017}, one needs to consider the subleading BMS diffeomorphisms and therefore the newly found subleading BMS charges at the subleading orders. Our idea is simple and the expectation is reasonable. At the leading order of the gravitational memory, if the leading BMS charge, namely the  Bondi mass aspect is conserved, then there is no flux at null infinity and therefore no displacement memory which we show it is not the case in the subleading order where we have a set of conserved charges. If we let the mass to vary, we have memory which is directly related to the change of the gravitational wave tensor, $\Delta C_{IJ}$ at null infinity. At the subleading order $r^{-3}$, we have a set of integrable and finite BMS charges which are related to some physical conserved charges. We are especially interested in the subleading gravitational memory if one considers newly found conserved charges at subleading orders. We will also use the method introduced in \cite{Bart:2020} to find the longitudinal memory in the $r$ direction.\\
 The paper is as follows: We begin by the asymptotically flat metric in four dimensions in section \ref{sec:Asymptotic metric}. Then we review the new method for the gravitational memory calculation based on null geodesics in section \ref{sec:GWMB}. The towers of gravitational memories is found in 
section \ref{sec:NUgauge}, where, we show the memory is not vanishing although having a set of conserved charges in some certain subleading orders. In section \ref{sec:longitudinal}, the longitudinal displacement is found and it is shown that this shift only shows itself in the subleading orders, mainly in order $1/r$. We conclude our work with some discussions in section \ref{discu}.

\section{Asymtotically 4.d flat metric in Bondi gauge}\label{sec:Asymptotic metric}
\subsection{Metric definition in the Bondi gauge}

The most general form of the metric in 4.d which is asymptotic to the flat metric in retarded Bondi gauge with coordinates $(u,r,x^I)$, $x^I=\{\theta,\phi\}$ takes the form

\begin{eqnarray}\label{m}
ds^2 =g_{\mu\nu}dx^\mu dx^\nu=-F e^{2\beta} \ du^2 -2 e^{2\beta} \ dudr+g_{IJ} (dx^I -C^I du)(dx^J -C^J du),
\end{eqnarray}

where $F$, $\beta$ and $C^I$ are some metric functions with the following fall-off conditions
\begin{eqnarray}\label{F}
&&F(u,r,x^I)=1+\sum_{n=0}^n \frac{F_n (u,x^I)}{r^{n+1}}, \ \ 
\b(u,r,x^I)=\sum_{n=0}^n \frac{\b_n (u,x^I)}{r^{n+2}}, \ \ C^J (u,r,x^I)=\sum_{n=0}^{n}\frac{C_n^J (u,x^I)}{r^{n+2}}.\nonumber\\
\end{eqnarray}
The inverse of the metric (\ref{m}) reads as
\begin{eqnarray}
g^{\mu\nu}=\begin{pmatrix}
 0 & -e^{-2\b} & 0\\
-e^{-2\b}& F e^{-2\b} & -e^{-2\b} C^J\\
  0&-e^{-2\b} C^I&g^{IJ}
 \end{pmatrix}.
\end{eqnarray}
The spatial part of the metric has the following fall-off at large-$r$ expansion defined in \cite{Godazgar:2018} as
\begin{eqnarray}\label{se}
g_{IJ}(u,r,x^I)=r^2 h_{IJ}=r^2 \gamma_{IJ}+rC_{IJ}(u,x^I)+\frac{C^2 \gamma_{IJ}}{4}+\frac{D_{IJ}(u,x^I)}{r}+\frac{E_{IJ}(u,x^I)}{r^2}+\mathcal{O}(r^{-2}).\nonumber\\
\end{eqnarray}
where, $\gamma_{IJ}$ is the metric on the 2-sphere, $C_{IJ}$ describes the gravitational waves and the square of  the so called ``Bondi news tensor'', $N_{IJ}=\p_u C_{IJ}$ is proportional to the flux of energy at the null boundary of the space time and $C^2=C_{IJ}C^{IJ}$. Fixing the gauge condition as follows 
\begin{eqnarray}\label{GC}
\frac{1}{r^2}\text{det}(g_{IJ})=\text{det}(\gamma_{IJ})=\sin^2\t,
\end{eqnarray}
keeps the spatial part of the metric topologically  spherical and gives us the following constraints about the functions $C_{IJ}$, $ D_{IJ}$ and $E_{IJ}$,

\begin{eqnarray}\label{traceh}
&&\frac{1}{r^2}\text{det}(g_{IJ})=\text{det}(\gamma_{IJ}+\frac{C_{IJ}}{r}+\frac{C^2 \gamma_{IJ}}{4r^2}+\frac{D_{IJ}}{r^3}+\frac{E_{IJ}}{r^4}+{\mathcal{O}(r^{-4})})\nonumber\\
&&=\text{det}(\gamma_{IJ}) \   [1+\frac{C_I^I}{r}+\frac{D_I^I}{r^3}+\frac{E_I^I}{r^4}-\frac{C^2 C_I^I}{4r^3}-\frac{C_{KL} D^{KL}}{r^4}{+}\frac{C^4 }{16 r^4}+..],\nonumber\\
\end{eqnarray}
which implies

\begin{eqnarray}\label{trace}
 \text{tr}C=0,  \ \ 
 \text{tr} D=0,  \ \ \text{tr} E=C_{IJ} D^{IJ}-\frac{C^4}{16}.
\end{eqnarray}
Assuming appropriate fall-off condition for energy-momentum tensor at each order, some of the metric functions in metric (\ref{m}) read as\footnote{See section (2.2) of \cite{Godazgar:2018} for the complete derivation of the metric functions in metric (\ref{m}).}
\begin{eqnarray}\label{coefficients}
&&  \partial_u F_0 = -\frac{1}{2} D_I D_J \partial_u C^{IJ} + \frac{1}{4} \partial_u C^{IJ} \partial_u C_{IJ},\\
&&\beta_0 = -\frac{1}{32}\, C^2,\ \ \beta_2 =\frac{1}{128}\, (C^2)^2 -\frac{3}{32} D_{IJ} C^{IJ}  ,\\
&& C_0^I = -\half D_J C^{IJ}.\label{C0}
\end{eqnarray}
One can also re-parametrise $h_{IJ}$, the spatial part of the metric (\ref{m}). The parametrisation of the metric is indeed very useful and crucial for calculating the coefficients in the metric functions described in (\ref{m}) and in all the calculations in the current paper. See Appendix \ref{param} for the parametrisation of the metric.
\section{Gravitational memory in the bulk}\label{sec:GWMB}
 The gravitational memory effect is defined as the permanent shift in the relative distance of a pair of inertial observers\footnote{See \cite{strominger:2016} for a detailed discussion on different types of observers.} stationed near $\mathcal{I}^+$ due to the burst of energy. The main purpose of this paper is to find the corrections to the gravitational memory away from null infinity which we call  ``subleading gravitational memory''. In fact, we aim to calculate the displacement memory in the bulk, close to the boundary of conformally compactified space time. One can use null geodesics instead of timelike geodesics and find the gravitational displacement \cite{Bart:2020}. It is shown that this method can be easily used for finding the gravitational memory effect in the bulk of the space time. We first briefly review this new method for the formulation of the gravitational memory in the bulk using ingoing null geodesics. One can measure the deviation between a pair of null geodesics $\eta^{\mu}|_v$, when one shoots light rays at time $v$. At some later time after the burst of the gravitational energy, another pair of light rays are considered and the deviation between the second pair of null geodesics is measured as $\eta^{\mu}|_{v^{\prime}}$. The new method for the formulation of the gravitational memory is based on comparing the geodesic deviation between two assumed pairs of null geodesics as follows
\begin{eqnarray}\label{deltaeta}
\Delta\eta^{\mu}:=\eta^{\mu}|_{v^{\prime}}-\eta^{\mu}|_{v},
\end{eqnarray} 
which is well-defined in any radius. $\eta^{\mu}$ is the deviation between a pair of null geodesics given by
\begin{eqnarray}\label{etaI}
\eta^{\mu}(r)=x^{\mu}(r)-x_0^{\mu}(r).
\end{eqnarray} 
$x_0^{\mu}(r)$ is  a geodesic generated by the vector $n_{\nu}=-\p_{\nu}  v$ and the neighbourhood geodesic $x^{\mu}(r)$ generated by small deformations $f$ given by
\begin{eqnarray}\label{geoo}
x^{\mu}(r)=\int^r g^{\mu\nu} (n_{\nu}-\partial_{\nu} f)+z^{\mu}.
\end{eqnarray}
$f$ is a function of the 2-sphere and $z^\mu$ indicates the ingoing location of the geodesic. 
In Eq. (\ref{etaI}), $r$ is an affine parameter. If we choose Newman-Unti (NU) coordinate, at a sufficiently large radius $r=r_L$, the worldlines $(v,r_L, x^I)$ are approximately inertial observers and the gravitational displacement due to the burst of energy between $v$ and $v^\prime$ can be computed easily by comparing the quantity $\eta^{\mu}$ before and after the flux at infinity. It is shown that this method is also applicable to find the gravitational memory in the bulk of the space time. We are in fact interested in the large-$r$ expansion of the gravitational memory which is near null infinity and in the bulk of the space time.
 \section{Subleading memory effect in Newman-Unti gauge} \label{sec:NUgauge}
The asymptotically flat space time metric can be written in NU coordinate $(v,r,x^I)$. In this gauge null foliations of space time are describe with parameter $v$ while $r$ is affine for generators of geodesics in the hypersurfaces $\Sigma_v$ of constant $v$.  Starting from the form of the metric (\ref{m}) in Bondi gauge with radial coordinate $r$, one can find the asymptotically flat metric in NU gauge by the following radial coordinate change according to Barnich and Lambert \cite{Barnich} as
\begin{eqnarray}
r_{\text{NU}}=r-\int_{r}^{\infty} dr^\prime \ (e^{2\b}-1)=r-\frac{2\b_0}{r}-\frac{2(\b_0^2+\b_2)}{3r^3}+{\mathcal{O}(r^{-3})}.
\end{eqnarray}
As we are interested in the spatial gravitational displacement $\Delta \eta^I$ at some fixed $r$, we need the spatial part of the metric in NU gauge. We will address the displacement in the longitudinal direction, $\Delta \eta^r$ as well later in this the paper. 
The spatial part of the asymptotically flat metric in NU gauge can be found as 
\begin{eqnarray}\label{UNguu}
g^{IJ{(\text{NU})}}=\frac{\gamma^{IJ}}{r_{\text{NU}}^2}+\frac{\mathcal{A}^{IJ}}{r_{\text{NU}}^3}+\frac{\mathcal{B}^{IJ}}{r_{\text{NU}}^4}+\frac{\mathcal{G}^{IJ}}{r_{\text{NU}}^5}+\frac{\mathcal{H}^{IJ}}{r_{\text{NU}}^6}+\mathcal{O}(r_{{\text{NU}}}^{-7}),
\end{eqnarray}
with
\begin{eqnarray}
&& \mathcal{A}^{IJ}=-C^{IJ}, \ \ \mathcal{B}^{IJ}=\frac{3}{8} C^2 \gamma^{IJ}, \ \ \mathcal{G}^{IJ}=-(D^{IJ}+\frac{3}{16} C^2 C^{IJ}),\nonumber\\
&& \mathcal{H}^{IJ}=-E^{IJ}+\frac{9}{8} (D^{KL} C_{KL}-\frac{1}{16} C^4)\ \gamma^{IJ}+\frac{5C^4}{64}\gamma^{IJ}.
\end{eqnarray}
We assume the following large-$r$ expansion for the spatial gravitational displacement\footnote{The spatial memory in Eq. (\ref{D}) starts from $n=2$ but that is just because of the factor $1/r^{2}$ in $g^{IJ}$. } as

\begin{eqnarray}\label{D}
\mathscr{D}^{I}=\sum_{n=2}^n \frac{\mathscr{D}^{I(n-2)}}{r^n}.
\end{eqnarray}
 Using Eqs. (\ref{deltaeta}) and ({\ref{UNguu}), one can find the the gravitational memory in each order. One should also consider the newly found  BMS chrages at subleading orders which are actually related to the conserved  Newman-Penrose charges.  Therefore, the change in the geodesic deviation with leading order deviation $d^I=\frac{1}{r} \gamma^{IJ} \p_J f$, gives us a shift which by considering the conserved charges at each order corresponds to the gravitational displacement at each order.\\
\subsection{Memory effect and the BMS charge at $\mathcal{O}(r^{0})$}
According to \cite{Barnich}, the variation of the leading BMS charge reads as
\begin{eqnarray}\label{I0}
\delta \mathcal{I}_0=\delta (-2s F_0)+\frac{s}{2} \p_u C_{IJ}\delta C^{IJ},
\end{eqnarray}
where $s$ is referred to the supertranslations and 
\begin{eqnarray}
\delta C^{IJ}=s \p_u C^{IJ}-2 D^I D^J s,
\end{eqnarray}
is the variation of $C^{IJ}$ under the action of supertranslation and $D^I$ is the covariant derivative with respect to $\gamma^{IJ}$. The first term in charge (\ref{I0}) is the integrable part of the charge and $F_0$ is proportional to the Bondi mass aspect $m_B$. The second term is the non-integrable part of the BMS charge and corresponds to the radiation at null infinity. If one sets the non-integrable part of charge to zero, then $\mathcal{I}_0$ is integrable and $F_0$ is conserved which means there is no radiation at null infinity and therefore $C^{IJ}$ is a free data at null infinity. But if we allow $F_0$ to vary, there will be a flux of gravitational radiation at null infinity and  $\p_u C^{IJ}$ is no longer zero. The gravitational memory at leading order and right at null infinity reads as 
\begin{eqnarray}
\Delta \eta^{I(0)}=-\frac{\p^J f}{2r_{NU}^2} \Delta C_J^I=-\frac{\p^J f}{2r^2} \Delta C_J^I.
\end{eqnarray}
Therefore,
\begin{eqnarray}
\mathscr{D}^{I(0)}=-\frac{d^J}{2r} \Delta C_J^I.
\end{eqnarray}
At this order one can also find the gravitational shift between a pair of timelike observers using the geodesic equation \cite{strominger:2016,strominger:2017}. See Appendix \ref{AppNP} for the derivation of the leading memory at null infinity in NP formalism using timelike geodesics.

\subsection{Memory effect and BMS charge at $\mathcal{O}(r^{-1})$}
At this order the BMS charge is identically zero due to the Einstein equations and strong enough choice for the fall off of the energy-momentum tensor, thus  $\mathcal{I}_1=0$. Therefore, there is no conserved quantity in this order and the gravitational shift at this order is 
\begin{eqnarray}
\Delta \eta^{I(1)}=-\frac{\p^I f}{8r_{NU}^3} \Delta (C^2)=-\frac{\p^I f}{8r^3} \Delta (C^2).
\end{eqnarray}
Therefore,
\begin{eqnarray}
\mathscr{D}^{I(1)}=-\frac{d^I}{8r^2} \Delta (C^2).
\end{eqnarray}
\subsection{Memory effect and BMS charge at $\mathcal{O}(r^{-2})$}
At this order the BMS charge is integrable as the non-integrable part of the charge is zero $\delta \mathcal{I}_2^{\text{(non-int)}}=0$, for $l=0$ or $l=1$ spherical harmonics with the constraint $D_I D_J s =\frac{1}{2} \gamma_{IJ} \Box s$ on the supertranslation function $s$. The integrable part of the charge is given by integrating 
\begin{eqnarray}
\delta \mathcal{I}_2=D_I D_Js\  (-D^{IJ}+\frac{1}{16} C^2 C^{IJ}).
\end{eqnarray}
Due to the trace free property of $C_{IJ}$ and $D_{IJ}$ defined in (\ref{trace}), the BMS charge at this order is zero and therefore the memory reads as
\begin{eqnarray}
\Delta \eta^{I(2)}=-\frac{\p^J f}{4r_{NU}^4}(\Delta ( D_J^{I}-\frac{1}{16} C^2 C_J^{I})+\frac{1}{4} \Delta (C^2 C_J^{I}))=-\frac{\p^J f}{4r^4}\Delta ( D_J^{I}-\frac{1}{16} C^2 C_J^{I}).
\end{eqnarray}
Thus,
\begin{eqnarray}\label{D}
\mathscr{D}^{I(2)}=-\frac{d^J}{4r^3}\Delta ( D_J^{I}-\frac{1}{16} C^2 C_J^{I}).
\end{eqnarray}
\subsection{Memory effect and BMS charge at $\mathcal{O}(r^{-3})$}
At this order the non-integrable part of the BMS charge is zero and if one considers the supertranslation $s$ to be an $l=2$ spherical harmonic, there is a new set of integrable charges which are directly related to the conserved NP charges as
 \begin{eqnarray}
 \mathcal{I}_3=s D_I D_J (-E^{IJ}+\frac{1}{2}(D^{KL} C_{KL}-\frac{1}{16} C^4)\gamma^{IJ}).
 \end{eqnarray}
 The above expression is directly related to the component $\Psi_0^1$ of the Weyl scalars\footnote{See Eq. (4.30) of \cite{Godazgar:2018}.}.  Using (\ref{deltaeta}), the gravitational shift is given by
 \begin{eqnarray}
&&\Delta\eta^{I(3)}=-\frac{\p_J f}{4r_{NU}^5}\Delta(E^{IJ} +\frac{9}{5} \Big(-E^{IJ}+\frac{1}{2} (D^{KL}C_{KL}-\frac{1}{16}C^4)\gamma^{IJ}\Big)+\frac{1}{16} C^4 \gamma^{IJ})\nonumber\\
&&=-\frac{\p_J f}{4r^5}\Delta(E^{IJ} +\frac{9}{5} \Big(-E^{IJ}+\frac{1}{2} (D^{KL}C_{KL}-\frac{1}{16}C^4)\gamma^{IJ}\Big)-\frac{1}{32} C^4 \gamma^{IJ}).
\end{eqnarray}
But due to the existence of set of integrable conserved charges at this order, the gravitational shift is obtained by
\begin{eqnarray}\label{E}
\mathscr{D}^{I(3)}=-\frac{d^J}{4r^4}\Delta (E_J^I-\frac{\delta_J^I}{32} C^4 ).
\end{eqnarray}
The above equation shows that even if the charge remains conserved in this order, still there is a non-trivial displacement. However, we just considered the subleading BMS charges which are related to the real part of the weyl scalars.

\section{Longitudinal displacement}\label{sec:longitudinal}
In addition to $\Delta \eta^I$, one can also consider the gravitational shift in the longitudinal direction as $\Delta \eta^r$. Using Eq. (\ref{deltaeta}), the shift on the $r$ direction can be read as
\begin{eqnarray}
\Delta \eta^r=-\int^r g^{rI} \p_I f.
\end{eqnarray}
Using Eq. (\ref{C0}), the above equation can be written as\footnote{The large-$r$ expansion of  $g^{rI}$ component of the metric reads as  
\begin{eqnarray}\label{bondigrI}
g^{rI}=-\frac{C_0^I}{r^2}-\frac{C_1^I}{r^3}-\frac{2\b_0 C_0^I+C_2^I}{r^4}+\mathcal{O}(r^{-5}).\nonumber
\end{eqnarray}
}
\begin{eqnarray}\label{long}
&&\Delta \eta^r=\frac{D^J\Delta C_{IJ}}{2r}\p^I f+\mathcal{O}(r^{-2}).\label{sr}
\end{eqnarray}
But vanishing of the curvature in the asymptotic region implies
\begin{eqnarray}
\Delta C_{IJ}=2 D_I D_J s,
\end{eqnarray}
Thus, Eq. (\ref{long}) shows the relative radius change between two null geodesics which is the same shift in $r$ direction derived in\cite{strominger:2016}\footnote{See Eq. (4.13) in the reference.} as $\delta r$ when one considers the action of the supertranslations $s$ on the initial separation vector between the detectors $\zeta$ as $\mathcal{L}_s \zeta$. Therefore, the shift in the $r$ direction between two observers can be neatly recast using the new method for finding the gravitational memory based on null geodesics.

\section{Discussion}\label{discu}
In this paper, we studied the non linear memory in the bulk of the space time around null infinity. Considering the large-$r$ expansion of the BMS conserved charges at each order, we showed that despite the existence of a set of integrable charges at order $r^{-3}$, the memory is not zero contrary to the case in the leading order. That may be related to the fact that the physical meaning of the NP charges is not very clear especially in the non-linearised theory of gravity while the interpretation of the charge in the leading order is the Bondi mass aspect. However, we only considered the supertranslation charges at subleading orders. Large-$r$ expansion of the superrotation charges which may be related to some conserved physical charges and the connection to the subleading memory remains to be explored in future works. However, the observable for the spin memory is a time delay \footnote{See \cite{Sabri,Mirzaiyan}  for the connection of the spin memory and superrotations.} while in this paper the displacement memory is considered. Moreover, the gravitational waves and therefore the infinite towers of memories are characterized by the asymptotic shear $\sigma^0$  and according to \cite{Mao:2020}, the subleading memories are defined by different choice of the asymptotic shear. However as it is obvious from Eqs. (\ref{D}) and (\ref{E}), the displacement memories are not characterized only by the shear\footnote{The tower of memories should be only dependent on the functions $f_0$ and $g_0$ (See Appendix \ref{param}). However, the subleading memory at order $r^{-2}$ and $r^{-3}$ also depends on the functions $f_2$, $g_2$, $f_3$ and $g_3$. Suppose in a physical system we let the Bondi mass be conserved. Therefore, there will be no gravitational displacement memory at null infinity and also not around the null boundary.} which is related to the flux at infinity. We may solve this in a future work by considering the tower of dual BMS charges \cite{HGodazgar:2018}. Also despite the leading order in which the memory is produced by the non-integrable part of the leading charge, the non-integrable parts of the subleading supertranslation charges do not play any role in the subleading memory effect. Finally, in addition to the standard memory, a displacement in the radial direction has been calculated. It is shown that the shift in the $r$ direction shows itself only away from null infinity and it is exactly the change in the relative radius between two detectors when the supertranslations act on the initial separation vector between two observers before passage of the gravitational waves. This confirms the method we used in the body of the paper led to consistent results with previous works.
\section*{Acknowledgements}
We are grateful of Maryam Aghaei, Henk Bart, Ajit Mehta and especially Hadi Godazgar for useful discussions. We thank Axel Kleinschmidt and Maarten van de Meent for the comments on the draft.

\appendix
\section{Parametrisation of the spatial part of the asymptotically flat metric}\label{param}
 Following \cite{BMS,Godazgar:2018}, one can do a parametrization of the spatial part of the metric (\ref{m}) as 
\begin{eqnarray}\label{smetric}
h_{IJ} dx^I dx^J=\frac{1}{2}(e^{2f} +e^{2g})d\theta^2 +  (e^{f-g} +e^{-f+g})\sin\theta \ d\t d\phi +\frac{1}{2}\sin^2 \theta \ (e^{-2f} +e^{-2g})d\phi^2.\nonumber\\
\end{eqnarray}
with $f(u,r,\theta,\phi)$ and $g(u,r,\theta,\phi)$ defined as 
\begin{eqnarray}\label{fg}
f(u,r,\theta,\phi)=\sum_{n=0}^\infty \frac{f_n}{r^{n+1}}, \ \ \ \ g(u,r,\theta,\phi)=\sum_{n=0}^\infty \frac{g_n}{r^{n+1}}.
\end{eqnarray}
with the constraint $f_1=g_1=0$, that comes from the Sommerfeld radiation condition\footnote{The condition is also called the regularity condition on the metric and it is the consistent behaviour of the metric functions at far distances. See section 4 of \cite{BMS} and especially Eq. (4.10) of \cite{sachs}.} which describes the behaviour of the field in asymptotic region as $r\rightarrow\infty$. One can expand the spatial metric (\ref{smetric}) in terms of $f_n$ and $g_n$ and find the tensors $C_{IJ}$, $D_{IJ}$ and $E_{IJ}$ in terms of $f$ and $g$ as

\begin{align}\label{Cfg}
&C_{00}=-\frac{1}{\sin^2\theta} C_{11}=f_0+g_0, \ \ C_{01}=C_{10}=(f_0-g_0) \sin\theta,\\
&D_{00}=-\frac{1}{\sin^2\theta} D_{11}=f_2+g_2+\frac{2}{3} (f_0^3+g_0^3),\nonumber\\
&D_{01}=D_{10}=(f_2-g_2+\frac{1}{6} (f_0-g_0)^3) \sin\theta,\\
&E_{00}=f_3+g_3+\frac{1}{3}f_0^4+2f_0 f_2+2g_0 g_2 +\frac{1}{3} g_0^4,\nonumber\\
&E_{11}=-(f_3+g_3-\frac{1}{3}(f_0^4+6f_0 f_2+6g_0g_2+g_0^4))\sin^2 \theta,\nonumber\\
&E_{01}=E_{10}=(f_3-g_3) \sin\theta.
\end{align}
The the above parametrisation is indeed essential for finding the metric functions defined in \cite{Godazgar:2018}.
\section{Memory effect at null infinity in Newman-Penrose formalism}\label{AppNP}
The gravitational memory can be also written in NP formalism\footnote{For the NP formalisation of the infinite towers of gravitational memory see \cite{Mao:2020}.}. The Newman-Penrose formalism begins with a choice of complex null frame $\{l,n,m,\bar{m}\}$. We begin with the asymptotic form of the metric as
\begin{eqnarray}
&&ds^2 =-F e^{2\beta} \ du^2 -2 e^{2\beta} \ dudr+r^2 h_{IJ} (dx^I -C^I du)(dx^J -C^J du)\nonumber\\
&&=-e^{2\beta} du \ (F du+2dr)+2 r^2 (\hat{m}_{(I} \bar{\hat{m}}_{J)})  (dx^I -C^I du)(dx^J -C^J du).
\end{eqnarray}
Using $g_{ab}=E_a^\mu E_b^\nu \eta_{\mu\nu}$, the orthonormal tetrad $\{E_\alpha\}$ reads as 
\begin{eqnarray}
&&E_0^\perp=e^{2\beta} du, \ \ E_1^\perp=\frac{1}{2}(F du+2 dr), \ \  E_2^\perp=r \hat{m}_I (dx^I-C^I du), \ \  E_3^\perp=r \bar{\hat{m}}_I (dx^I-C^I du).\nonumber\\
\end{eqnarray}
Accordingly, a complex null frame can be introduced as
\begin{eqnarray}
&&l=\p_r, \ \ n=e^{-2\beta} [\p_u-\frac{1}{2} F \p_r+C^I \p_I], \ \ m=\frac{\hat{m}^I}{r}\p_I, \ \ \bar{m}=\frac{\bar{\hat{m}}^I}{r}\p_I.
\end{eqnarray}
The scalar products of the tetrad vectors vanishes apart from
\begin{eqnarray}
l^\mu n_\mu=
-m^\mu \bar{m}_\mu=-1.
\end{eqnarray}
The coordinate basis can be read as
\begin{eqnarray}
\p_u=e^{2\beta} n +\frac{1}{2} F l -r h_{IJ}C^I \ (m \bar{\hat{m}}^J+\bar{m} \hat{m}^J), \ \ \p_r=l, \ \ \p_I=r h_{IJ} \ (m \bar{\hat{m}}^J+\bar{m} \hat{m}^J).\nonumber\\
\end{eqnarray}
 We will use the above basis for writing the Riemann tensor in terms of the Weyl scalars.
In general the geodesic equation shows how the inertial particles undergo a mutual acceleration due to the curvature of the space time. The relative acceleration between two time like geodesics is descibed by the equation of geodesic as follows
\begin{eqnarray}\label{geo}
\frac{D^2 \xi^\alpha}{du^2}=-R_{\beta \gamma \delta}^{\alpha} v^{\beta}v^{\delta} \xi^{\gamma},
\end{eqnarray}
where, $v^\alpha=(1,0,0,0)$ is the four vector velocity  of  a timelike observer and $\xi^\alpha$ is the deviation vector between two neighbouring geodesics. \\
We now find the large-$r$ expansion of the gravitational displacement due to the passage of the gravitational  burst. We consider the following large-r expansion for the $R_{\beta \gamma \delta}^{\alpha}$ and the four-velocity of the inertial observer\footnote{One may consider the corrections to the four-velocity of the observer (See exercise 13 of \cite{strominger:2017}), but actually the only important component of the four-vector velocity of the inertial observer is $v^{u(0)}$ to remain inertial during the gravitational energy burst.} as
 \begin{eqnarray}
R_{\beta \gamma \delta}^{\alpha}=\sum_{n=1}^n \frac{ R_{\beta \gamma \delta}^{\alpha(n)}}{r^n}, \ \ \ \ v^{\alpha}=\sum_{m=0}^m \frac{v^{\alpha(m)}}{r^m}.
 \end{eqnarray}
The Riemann tensor component which is interesting for the leading memory reads as

\begin{eqnarray}\label{ruiuj}
R_{uIuJ}=R_{\mu\nu\rho\sigma} (\p_u)^\mu (\p_u)^\rho (\p_I)^\nu (\p_J)^\sigma.
\end{eqnarray}
In vacuum $R_{\mu\nu}=0$ and the Riemann tensor becomes as same as the Weyl tensor. Of course, that is not the general case and one may investigate the problem in the presence of matter. So Eq. (\ref{ruiuj}) can be written as 
\begin{eqnarray}
R_{uIuJ}=C_{\mu\nu\rho\sigma} (\p_u)^\mu (\p_u)^\rho (\p_I)^\nu (\p_J)^\sigma.
\end{eqnarray}
Five complex Weyl scalars\footnote{See \cite{Szekeres} for the physical interpretation of the Weyl scalars in asymptotic region at large distance from the source. In particular, $\Psi_4$ describes the gravitational radiation at null infinity. This is also clear from the non-integrable part of the leading charge (\ref{0charge}) as $\Psi_4=- \bar{\ddot{\sigma}}^0=-\ddot{h}_+ +i \ddot{h}_{\times}$. } are as follows
\begin{eqnarray}\label{NPcharges}
&&\Psi_0= l^a m^b l^c m^d C_{abcd}, \ \ \Psi_1=l^a n^b l^c m^d C_{abcd}, \ \ \Psi_2=l^a m^b \bar{m}^c n^d C_{abcd},\nonumber\\
&& \ \ \Psi_3=l^a n^b \bar{m}^c n^d C_{abcd},  \ \ \Psi_4=n^a \bar{m}^b n^c \bar{m}^d C_{abcd}.
\end{eqnarray}
Assuming
\begin{eqnarray}
\p_u^\mu=e^{2\b}n^\mu +\frac{1}{2} F l^\mu -r h_{KL} C^K (m^\mu \bar{\hat{m}}^L+\bar{m}^\mu \hat{m}^L),\ \  \p_I^\nu=rh_{IN} (m^\nu \bar{\hat{m}}^N+\bar{m}^\nu \hat{m}^N).\nonumber\\
\end{eqnarray}
After a long but straightforward calculation, the $R_{uIuJ}$ component of the Riemann tensor can be written in terms of Newman-Penrose charges as 
\begin{align}\label{Ruiuj}
R_{uIuJ}&=\sum_{n=0}^4 C_n \Psi_n+\sum_{n=0}^4 C_n^\prime \bar{\Psi}_n\nonumber\\
&=C_0 \Psi_0+C_{0}^\prime \bar{\Psi}_0+C_1 \Psi_1+C_{1}^\prime \bar{\Psi}_1+C_2 \Psi_2+C_{2}^\prime \bar{\Psi}_2+C_3 \Psi_3+C_{3}^\prime \bar{\Psi}_3+C_4 \Psi_4+C_{4}^\prime \bar{\Psi}_4.\nonumber\\
\end{align}
For $R_{uIuJ}$ to be real-valued, $C_n=\bar{C_n^\prime}$. The $C_n$s coefficients read as 

\begin{align}
C_0&=\bar{C}_0^\prime=\sum_{n=-2}^n\frac{C_0^{(n)}}{r^n}=\frac{1}{4} r^2 F^2 h_{IN} h_{JP} \bar{\hat{m}}^N \bar{\hat{m}}^P,\\
C_1&=\bar{C}_1^\prime=\sum_{n=-1}^n\frac{C_1^{(n)}}{r^n}=\frac{1}{2} r^3 F h_{IJ} h_{KL} C^K  \bar{\hat{m}}^L-r^3 F h_{IN}h_{JP} h_{KL} C^K \bar{\hat{m}}^N \bar{\hat{m}}^P \hat{m}^L,\\
C_2&=\bar{C}_2^\prime=\sum_{n=-2}^n\frac{C_2^{(n)}}{r^n}=-\frac{1}{2} r^2 e^{2\beta} F  h_{IJ}+r^4 h_{IN}h_{JP}h_{KL}h_{RS} C^R C^K \hat{m}^N \hat{m}^P \bar{\hat{m}}^L \bar{\hat{m}}^S\nonumber\\
&+r^4 h_{IN}h_{JP}h_{KL}h_{RS} C^K C^R \bar{\hat{m}}^N \bar{\hat{m}}^P \hat{m}^L \hat{m}^S-r^4 h_{IN}h_{JP}h_{KL} C^K C^L \bar{\hat{m}}^N,\\
C_3&=\bar{C}_3^\prime=\sum_{n=-1}^n\frac{C_3^{(n)}}{r^n}=-2r^3 e^{2\b} h_{IN} h_{JP} h_{KL} C^K \bar{\hat{m}}^N \bar{\hat{m}}^P \hat{m}^L+r^3 e^{2\b} h_{IJ} C^K  \bar{\hat{m}}^L,\\
C_4&=\bar{C}_4^\prime=\sum_{n=-2}^n\frac{C_4^{(n)}}{r^n}=r^2 e^{4\beta} h_{IN}h_{JP} \hat{m}^N \hat{m}^P \hat{m}^P.
\end{align}
In order to write the large-$r$ expansion of $R_{uIuJ}$, we use the peeling property of the Weyl tensor and find the 
\begin{eqnarray}
&&\Psi_0=\sum_{n=0}^n \frac{\Psi_0^n}{r^{5+n}},\ \ \Psi_1=\sum_{n=0}^n \frac{\Psi_1^n}{r^{4+n}},\ \ 
\Psi_2=\sum_{n=0}^n \frac{\Psi_2^n}{r^{3+n}},\ \ \Psi_3=\sum_{n=0}^n \frac{\Psi_3^n}{r^{2+n}},\ \
\Psi_4=\sum_{n=0}^n \frac{\Psi_4^n}{r^{1+n}}.\nonumber\\
\end{eqnarray}
At $\mathcal{O}(r)$ the component of the Riemann tensor reads as
\begin{eqnarray}
&&R_{uIuJ}^{(-1)}=C_4^{(-2)} \Psi_4^0 +C_4^{\prime(-2)} \bar{\Psi}_4^0=-\gamma_{IN}\gamma_{JP} (\hat{m}^N \hat{m}^P \bar{\ddot{\sigma}}^0 +\bar{\hat{m}}^N \bar{\hat{m}}^P {\ddot{\sigma}}^0)\nonumber,
\end{eqnarray}
where, the asymptotic shear $\sigma^0$ is
\begin{eqnarray}\label{sigma}
\sigma^0=\frac{1+i}{2}(f_0+i g_0).
\end{eqnarray}
 $\hat{m}^1$ and $\hat{m}^2$ at the leading order should be defined such that $h^{IJ}=\hat{m}^I \bar{\hat{m}}^J+\hat{m}^J \bar{\hat{m}}^I$. Therefore, we define
 \begin{eqnarray}\label{mmb}
 \hat{m}^1=\frac{(1+i)}{2}, \ \ \ \ \ \  \hat{m}^2=\frac{(1-i)}{2 \sin\t}.
 \end{eqnarray}
 Using Eqs. (\ref{sigma}) and (\ref{mmb}), (\ref{Ruiuj}) can be written as
 
 \begin{align}\label{lm}
 R_{u\t u\t}^{(-1)}&=-\gamma_{\t\t}\gamma_{\t\t} (\hat{m}^1 \hat{m}^1 \bar{\ddot{\sigma}}^0 +\bar{\hat{m}}^1 \bar{\hat{m}}^1 {\ddot{\sigma}}^0)=-\frac{1}{2} (\p_u^2 f_0+\p_u^2 g_0)=-\frac{1}{2} \p_u^2 C_{\t\t},\\
 R_{u\phi u\phi}^{(-1)}&=-\gamma_{\phi\phi}\gamma_{\phi\phi} (\hat{m}^2 \hat{m}^2 \bar{\ddot{\sigma}}^0 +\bar{\hat{m}}^2 \bar{\hat{m}}^2 {\ddot{\sigma}}^0)=\frac{1}{2} \sin^2\t \ (\p_u^2 f_0+\p_u^2 g_0)=-\frac{1}{2} \p_u^2 C_{\phi\phi},\\
   R_{u\t u\phi}^{(-1)}&=-\gamma_{\t\t}\gamma_{\phi\phi} (\hat{m}^1 \hat{m}^2 \bar{\ddot{\sigma}}^0 +\bar{\hat{m}}^1 \bar{\hat{m}}^2 {\ddot{\sigma}}^0)=-\frac{1}{2} \sin\t (\p_u^2f_0-\p_u^2g_0)=-\frac{1}{2}\p_u^2 C_{\t\phi}.
\end{align}
 Therefore, at the leading order, $R_{uIuJ}$ is found to be
  \begin{eqnarray}
&& R_{uI uJ}^{(-1)}=-\frac{1}{2}\p_u^2 C_{IJ}.
 \end{eqnarray}
 And the gravitational displacement reads as the standard gravitational memory displacement in \cite{strominger:2016}. At this order the standard BMS charge in terms of NP quantities is defined as \cite{Godazgar:2018}
 \begin{eqnarray}\label{0charge}
 \mathcal{Q}_0=-\frac{1}{2\pi G}\int d\Omega Y_{lm} \mathfrak{R}(\Psi_2^0+\sigma^0\p_u\bar{\sigma}^0).
 \end{eqnarray}
 If $\Psi_2^0$ is conserved, then $\p_u \Psi_2^0=0$ and as 
 \begin{eqnarray}
\psi_2^0 - \bar{\psi}_2^0 = \bar{\sigma}^0 \partial_u \sigma^0 -  \sigma^0 \partial_u \bar{\sigma}^0 + \bar{\eth}^2 \sigma^0 - \eth^2 \bar{\sigma}^0,
 \end{eqnarray}
 in order of $\Psi_2^0$ to be conserved, $\sigma^0$ should be a free data at null infinity $\dot{\sigma^0}=0$. In this case, we can see from (\ref{lm}) that gravitational displacement is zero due to the existence of a conserved charge at this order. If $\Psi_2^0$ is not conserved then there is a shift in the relative distance of two inertial observer which is related to the non-integrable part of the charge (\ref{0charge}).

\end{document}